\documentstyle[prl,aps,epsf,twocolumn]{revtex}
\input epsf

\begin{document}
\draft
\twocolumn[\hsize\textwidth\columnwidth\hsize\csname@twocolumnfalse%
\endcsname

\preprint{}
\title{Weak localization of disordered quasiparticles in the mixed
  superconducting state}
\author{R.~Bundschuh$^{1}$, C.~Cassanello$^{2}$, D.~Serban$^{3}$ and
  M.R.~Zirnbauer$^{2}$}
\address{
${}^{1}${\it Department of Physics, UCSD, La Jolla, CA 92093-0319, 
U.S.A.} \\
${}^{2}${\it Institut f\"ur Theoretische Physik,
Universit\"at zu K\"oln, Z\"ulpicher Str. 77, D-50937, K\"oln, 
Germany} \\
${}^{3}${\it Service de Physique Th\'eorique, 
CE-Saclay, F-91191 Gif-Sur-Yvette, France.}
}

\date{September 24, 1998}
\maketitle

\begin{abstract}
  Starting from a random matrix model, we construct the low-energy
  effective field theory for the noninteracting gas of quasiparticles
  of a disordered superconductor in the mixed state.  The theory is a
  nonlinear $\sigma$ model, with the order parameter field being a
  supermatrix whose form is determined solely on symmetry grounds.
  The weak localization correction to the field-axis thermal
  conductivity is computed for a dilute array of $s$-wave vortices
  near the lower critical field $H_{{\rm c}1}$. We propose that weak
  localization effects, cut off at low temperatures by the Zeeman
  splitting, are responsible for the field dependence of the thermal
  conductivity seen in recent high-${T_{\rm c}}$ experiments by Aubin
  {\it et al.}
\end{abstract}
\vspace{0.5cm}

%\bigskip{PACS numbers: 74.40.+k, 74.25.Fy, 72.80.Ng, 73.23.-b}%}
]
\narrowtext

\section{Introduction}

The long wave length physics of phases of matter with spontaneously
broken symmetries is commonly described by an effective field theory
for the relevant order parameter.  For the problem of localization and
transport in disordered metallic systems at low temperature, the
appropriate ``order parameter'' is known \cite{efetov} to be a
supermatrix (or a matrix of dimension zero if the replica trick is
used), conventionally denoted by $Q$.  Three universality classes,
differing by their behavior under time reversal and spin rotations,
are widely known \cite{dyson}.  They are labeled by an index $\beta =
1, 2, 4$, and are traditionally referred to as the classes with
orthogonal, unitary, and symplectic symmetry.  We denote them by
$A{\rm I}$, $A$, and $A{\rm II}$ for short \cite{az_ns}.  In each
case, the field theory for $Q$ belongs to the general family of
nonlinear $\sigma$ models.  The field $Q$ contains the Goldstone modes
of a hidden symmetry \cite{wegner} connecting retarded and advanced
single-electron Green functions, which is broken by a nonzero density
of states.  At tree level one recovers the classical diffusion
approximation, which neglects quantum interference corrections due to
electron paths with loops.  Anharmonic terms in the field theory
represent ``interactions'' between the diffusion modes, giving rise to
so-called weak localization corrections to diffusion.  For the classes
$A{\rm I}$ and $A$ in dimension $d \le 2$, these interactions become
strong at large distance scales and thus cause localization of all
states, regardless of the strength of the disorder.

In recent years it was found that the $\beta = 1, 2, 4$ classification
is not exhaustive: for systems with symmetries of the particle-hole
(ph) type, the invariance group of the order parameter field $Q$
becomes {\it enlarged} in the vicinity of the ph-symmetric point.  One
instance of such symmetry enhancement are Dirac fermions in a random
gauge field \cite{gade,jjmv}, another are disordered quasiparticles
that exchange charge (but no energy) with a superconducting condensate
\cite{az_ns}.  Such systems exhibit novel spectral statistics and
transport properties.  Seven symmetry-enhanced universality classes
have been identified, and the corresponding order parameters $Q$ were
constructed from their random matrix limit in \cite{mrz_rss}.  The
main message is that $Q$ lives on a symmetric space -- in precise
technical language: on a Riemannian symmetric superspace -- in all
cases.  The nonlinear $\sigma$ models defined over such spaces are
known \cite{friedan} to be {\it attractive under the flow of the
  renormalization group}.  Therefore the order parameter of a given
disordered single-particle system, and in most cases its low-energy
effective field theory as well, can be inferred quite simply by
investigating the ergodic (or random matrix) limit.

The present report focusses on class $C$, which emerges for
noninteracting low-energy quasiparticles in a magnetic field and in
contact with a spin-singlet superconductor.  The defining condition
\cite{az_ns} is that the quasiparticle Hamiltonian be invariant under
${\rm SU}(2)$ rotations of the electron spin, whereas time reversal
invariance has to be broken.  Since a superconductor screens magnetic
fields, this universality class can only be realized in an {\it
  inhomogeneous} superconducting state -- unless time reversal
invariance is broken spontaneously.  In the very recent literature the
following realizations have appeared \cite{ks}: i) a metallic quantum
dot in the form of a chaotic billiard, subject to a magnetic flux and
bordering on a superconductor \cite{az_andreev,melsen}; ii)
quasiparticles in the core of an isolated vortex in a disordered 
$s$-wave superconductor \cite{bcsz}; and iii) a (quantum) disordered
version of a $d_{x^2-y^2}$ superconductor with orbital coupling to a
magnetic field \cite{mpafetal}.  The hallmark of class $C$ is that, in
contrast with the metallic class $A$, the weak localization correction
does {\it not} vanish \cite{bb}, in spite of the presence of a
magnetic field.  The persistence of weak localization in a field is
caused by nonstandard modes of quantum interference that appear when
impurity and Andreev scattering are simultaneously present.  In a
semiclassical picture, the effect can be understood as being due
\cite{az_ns} to quasiparticle {\it paths in which a loop is circled
  twice}, with the charge states during the first and second looping
being exactly opposite to each other.

To identify the order parameter field $Q$ and its low-energy effective
theory for class $C$, one may proceed in several ways.  The direct
method, due to \cite{ast} and worked out in detail for isolated
vortices of an $s$-wave superconductor in Ref.~\cite{bcsz}, is to
start from the BCS mean field Hamiltonian for the quasiparticles, set
up a supersymmetric generating functional for the Gorkov Green
function, introduce a composite field $Q$ to decouple the 4-vertices
produced by averaging over the disorder, integrate out the
quasiparticle fields, solve two saddle point equations for $Q$ in
sequence (the second of which turns out to coincide with the Usadel
equation \cite{usadel}), and finally expand in gradients of $Q$ to
obtain the low-energy effective theory.  The field theory so obtained
is a nonlinear $\sigma$ model, with $Q$ taking values in a Riemannian
symmetric superspace of type $D{\rm III}|C{\rm I}$, in agreement with
the random matrix analysis of \cite{mrz_rss}.  Its coupling constant
has the universal meaning of a conductivity for the conserved
probability (or energy) current transported by the quasiparticles.
Because quasiparticles also carry spin, the coupling constant may be
reinterpreted \cite{mpafetal} as a spin conductivity in the present
context.  (The latter interpretation fails for systems with spin-orbit
scattering or magnetic impurities, where spin is not conserved.)

Given the proper identification of the order parameter field $Q$, a
few qualitative conclusions are {\it immediate}.  According to the
renormalization theory of nonlinear $\sigma$ models \cite{friedan},
the sign of the one-loop renormalization group beta function in two
space dimensions is completely determined by the sign of the (Ricci)
curvature tensor relative to the metric tensor.  Since the curvature
of the Riemannian symmetric superspace of type $D{\rm III}|C{\rm I}$
is positive \cite{helgason}, the weakly coupled two-dimensional theory
renormalizes by logarithmic corrections towards strong coupling ({\it
  i.e.} strong disorder), which ultimately leads to localization of
all quasiparticle states at $T = 0$.  This localized phase was called
a ``spin insulator'' in \cite{mpafetal}.  In dimension $d = 1$ the
same corrections are present, but with a linear dependence on the
cutoff length.  In $d = 3$ the theory supports a delocalization
transition to a phase of extended states, the ``spin metal''
\cite{mpafetal}.  The addition of random classical Heisenberg impurity
spins (at subcritical concentration, so as to maintain
superconductivity) causes crossover from class $C$ to class $D$
\cite{az_ns}, with the nonlinear $\sigma$ model changing to type
$C{\rm I}|D{\rm III}$ \cite{mrz_rss}.  In the process, the sign of the
symmetric space curvature gets reversed, whence weak localization
turns into weak {\it anti\/}localization, making it possible for
extended states to exist {\it already in two dimensions}.

Our goal here is to extend the treatment of \cite{bcsz} and illustrate
some of the above general facts at the thermal transport of the class
$C$ quasiparticles of a disordered $s$- or $d$-wave superconductor in
the mixed state \cite{vortices}.  We assume the presence of
(nondescript) nonmagnetic impurities, which disorder the vortex array
and cause elastic scattering of the quasiparticles.  To tackle this
problem we will use a coarse grained or random matrix type of
approach, placing the emphasis on {\it symmetry considerations}.

\section{Effective field theory from an $N$-orbital model}

We begin our treatment by partitioning the superconductor into cells
of equal size, with each cell containing one vortex segment with a
length of the order of the elastic mean free path $\ell$.  Within each
cell we introduce (in the spirit of the real-space renormalization
group) a basis of $N$ quasiparticle wavefunctions that comprise the
relevant low-energy configurations.  The matrix of the Hamiltonian in
such a basis assumes a sparse block structure, with one block on the
diagonal for each cell, and with off-diagonal blocks that couple
neighboring cells.  If $i$ labels the cells and $a = 1, ..., N$ the
orbitals inside a cell, the ``coarse grained'' Hamiltonian is of the
form
\begin{eqnarray*}
  H &=& \sum_{\langle i , j \rangle} \sum_{ab} \Big( h_{ia,jb} 
  ( c_{ia\uparrow}^\dagger c_{jb\uparrow}^{\vphantom{\dagger}} + 
  c_{ia\downarrow}^\dagger c_{jb\downarrow}^{\vphantom{\dagger}} ) 
  \\
  &&+ \Delta_{ia,jb} 
  ( c_{ia\uparrow}^\dagger c_{jb\downarrow}^\dagger
  - c_{ia\downarrow}^\dagger c_{jb\uparrow}^\dagger ) / 2 + 
  {\rm h.c.} \Big) \;,
\end{eqnarray*}
where the sum over $i,j$ is restricted to $i = j$ and pairs of
neighboring cells.  The spin-singlet nature $(\uparrow \downarrow -
\downarrow \uparrow)$ of the coupling to the pairing field is dictated
by conservation of spin.  Fermi statistics then requires the complex
matrix $\Delta$ to be symmetric: $\Delta_{ia,jb} = \Delta_{jb,ia}$
\cite{bahcall}.  If we temporarily suppress the cell and orbital
indices, $H$ can be written in the schematic form $H = {\rm Tr}({\cal
  H} \tilde{\bf c} {\bf c}) + {\rm const}$, where
  $$ 
  \tilde{\bf c} = \pmatrix{ c_\uparrow^\dagger \cr c_\downarrow}
  \;, \quad
  {\bf c} = ( c_\uparrow^{\vphantom{\dagger}} \, c_\downarrow^\dagger )
  \;, \quad
  {\cal H} = \pmatrix{h^{\rm T} &\Delta^\dagger\cr \Delta &-h \cr} \;.
  $$
The symmetries of the Hamiltonian matrix ${\cal H}$ are summarized by
the equation ${\cal H} = - {\cal C} {\cal H}^{\rm T} {\cal C}^{-1}$,
with ${\cal C}$ being the symplectic unit ${\cal C} = i \sigma_2
\otimes {\bf 1}$.  Note that when the Zeeman energy $H_{\rm Z} = \mu B
\sum_{ia} (c_{ia\uparrow}^\dagger c_{ia\uparrow}^{\vphantom{\dagger}}
- c_{ia \downarrow}^\dagger c_{ia\downarrow}^{\vphantom{\dagger}})/2$
is taken into account, the ${\rm SU}(2)$ spin rotation invariance of
$H$ is broken down to a ${\rm U}(1)$ symmetry.

Disorder in the microscopic Hamiltonian gives rise to randomness in
${\cal H}$.  Because the universal properties at long wave lengths are
insensitive to the microscopic details, we have considerable freedom
in choosing the random Hamiltonian ${\cal H}$.  The simplest choice is
an $N$-orbital model with locally gauge-invariant disorder of the type
invented by Wegner \cite{n_orbital} for the purpose of describing the
universal physics of the Anderson localization transition for $\beta =
1, 2, 4$.  The crucial new feature in the present case is the relation
${\cal H} = - {\cal C} {\cal H}^{\rm T} {\cal C}^{-1}$, which is
invariant under symplectic transformations ${\cal H} \mapsto S {\cal
  H} S^{-1}$, $S^{\rm T} {\cal C} S = {\cal C}$.  We therefore adopt a
model with local ${\rm Sp}(2N)$ gauge invariance: the elements of the
matrix ${\cal H}$ are taken to be Gaussian distributed uncorrelated
random variables with zero mean, $\langle {\cal H} \rangle = 0$, and
second moments specified by
  $$
  \langle {\rm Tr}A{\cal H}_{ij} \, {\rm Tr}B{\cal H}_{kl}\rangle 
  = {w_{ij} \over 2N} \left( \delta_{ij}^{lk} {\rm Tr}AB - 
    \delta_{ij}^{kl} {\rm Tr} A{\cal C}B^{\rm T} {\cal C}^{-1} \right)
  $$
where $\delta_{ij}^{kl} = \delta_{ik} \delta_{jl}$, and $w_{ij}$ is a
rapidly decreasing function of the distance between the cells $i$ and
$j$.  Aside from respecting the symmetries and locality of the
Hamiltonian, this choice has the virtue of maximizing the information
entropy. The main benefit from using such a maximum entropy model is
that the introduction of the supermatrix $Q$, usually a tricky step
that requires some expertise, becomes straightforward as we now
proceed to show.

We replace the operators $\tilde{\bf c},{\bf c}$ by classical fields
$\tilde\psi,\psi$: $\tilde{\bf c}_{ia\alpha} {\bf c}_{jb\beta} \to
\sum_\sigma\tilde\psi_{ia\alpha,\sigma} \psi_{\sigma,jb\beta}$ and
integrate bilinears in $\tilde\psi, \psi$ against $\exp i{\rm
  Tr}({\cal H}-E)\tilde\psi\psi$ in the usual way to generate the
Gorkov Green function at energy $E$.  The introduction of a bosonic
partner ($\sigma = {\rm B}$) for each fermionic field ($\sigma = {\rm
  F}$) serves to cancel vacuum graphs by the mechanism of
supersymmetry.  There is one complication, however: the matrix
$\tilde\psi\psi$ does not share the symplectic symmetry of the
Hamiltonian.  To remedy this mismatch, we introduce an extra quantum
number (``pseudo charge'') $c = \pm 1$, so that the quasiparticle
fields expand to tensors $\tilde \psi_{ia\alpha, \sigma c}$ and
$\psi_{\sigma c, ia\alpha}$ \cite{mrz_rss}.  On imposing the
conditions $\tilde\psi = {\cal C}\psi^{\rm T} \gamma^{-1}$ and $\psi =
- \gamma\tilde\psi^{\rm T} {\cal C}^{-1}$, where $\gamma$ is a real
orthogonal matrix that will be specified shortly, we have the symmetry
$\tilde\psi\psi = - {\cal C} (\tilde\psi\psi)^{\rm T} {\cal C}^{-1}$
as desired.

Since the order parameter $Q$ is a local field, its nature can be
uncovered by looking at the Hamiltonian truncated to a single cell.
With this truncation temporarily in force, we introduce $Q$ as
follows:
\begin{eqnarray*}
  &&\int d{\cal H} \exp ( - N {\rm Tr}{\cal H}^2 / 2w_0
  + i {\rm Tr}{\cal H}\tilde\psi\psi ) \\
  &=& \int dQ \exp ( - N{\rm STr} Q^2 / 2w_0
  + i {\rm STr} Q \psi\tilde\psi ) \;.
\end{eqnarray*}
The equality is verified by using the cyclic invariance of the
(super)trace: ${\rm Tr}(\tilde\psi\psi)^2 = {\rm STr} (\psi \tilde
\psi)^2$.  The Hubbard-Stratonovitch field $Q$ is a $4 \times 4$
supermatrix which, by its coupling to $\psi\tilde\psi$, inherits the
symmetry
\begin{equation}
  Q = - \gamma Q^{\rm T} \gamma^{-1} \;.
  \label{osp}
\end{equation}
The constraints relating $\tilde\psi$ and $\psi$ to one another are
compatible only if $\gamma^2$ equals the superparity matrix ($+1$ on
bosons, $-1$ on fermions).  To meet this condition we put 
  $$
  \gamma = E_{\rm BB} \otimes \sigma_1 + E_{\rm FF} \otimes i\sigma_2 
  = \pmatrix{ \sigma_1 &0\cr 0 &i\sigma_2} \;.
  $$
Relation (\ref{osp}) is the defining equation of an orthosymplectic
Lie algebra and is invariant under $Q \mapsto T Q T^{-1}$ with $\gamma
(T^{-1})^{\rm T} \gamma^{-1} = T \in {\rm OSp}(2|2)$.  In the general
case, where Green functions at $n$ different energies are to be
averaged, $Q$ acquires matrix dimension $4n \times 4n$, and the
symmetry group gets enlarged to ${\rm OSp} (2n|2n) \equiv G$.

Returning now to the full lattice problem, introducing $Q_i$ for every
cell $i$ and integrating over the quasiparticle fields $\psi, \tilde
\psi$ we arrive at the following action functional:
  $$
  S / N = \sum_{ij} {w^{-1}}_{ij} {\rm STr} Q_{i} Q_{j} / 2 + 
  \sum_{i} {\rm STr} \ln (Q_i - \omega \otimes \Sigma_3)
  $$
where $(\Sigma_3)_{\sigma c,\sigma^\prime c^\prime} = \delta_{\sigma
  \sigma^\prime} (\sigma_3)_{c c^\prime}$ and $\omega$ is a diagonal
matrix containing the energies at which the quasiparticle Green
functions are to be evaluated.  Variation of $S$ yields the saddle
point equation $\sum_{j} {w^{-1}}_{ij} Q_{j} = (\omega\Sigma_3 -
Q_{i})^{-1}$, whose physical solution (dictated by causality of the
Green function) at $\omega = 0$ and homogeneous in space is $Q^0 = i v
\Sigma_3$ with $v^{-2} = \sum_{j} {w^{-1}}_{ij}$.  Low-energy
fluctuations result from setting $Q_{i} = T_{i} Q^0 T_{i}^{-1}$ and
taking $T_{i} \in G$ to vary slowly with the position of the cell $i$.
By expanding in gradients, the low-energy effective action for such
configurations at $\omega = 0$ is easily seen to be a nonlinear
$\sigma$ model, 
\begin{equation}
  S_0 = - {\pi\nu \over 8}\int\limits d^3x \, 
  {\rm STr} \Big( D_{\perp} (\nabla_\perp Q)^2 +
  D_{\parallel}(\nabla_\parallel Q)^2 \Big) \;,
  \label{sigmod0}
\end{equation}
where we have switched to continuous coordinates $x_\perp = (x,y)$ and
$x_\parallel = z$.  The parameter $\nu$ is the density of states of
the superconductor, and $D_\parallel , D_\perp$ are the field-axis and
transverse {\it effective} diffusion constants of the quasiparticle
gas.  We are using units $\hbar = 1$.  At finite $\omega$, the
field theory action is perturbed by a term 
\begin{equation}
  S_\omega = {i\pi\nu \over 2} \int d^3x \, 
  {\rm STr} \omega \Sigma_3 Q \;, \qquad S = S_0 + S_\omega \;.
\label{sigmod1}
\end{equation}
We have rescaled the field to $Q = T \Sigma_3 T^{-1}$.  Since this
expression for $Q$ is invariant under $T \to T k$ for $k = \Sigma_3 k
\Sigma_3 \in {\rm GL}(n|n) \equiv K$, the supermatrix $Q$ lives on a
coset space $G/K$.  If we parametrize $Q$ by 
  $$
  Q = \exp \pmatrix{0 &X\cr \tilde X &0\cr} \Sigma_3 \;,
  $$
positivity of $S_0$ or equivalently, stability of the functional
integral, requires $\tilde X_{\rm BB} = + X_{\rm BB}^ \dagger$ and
$\tilde X_{\rm FF} = - X_{\rm FF}^ \dagger$.  In invariant
mathematical language, this means $Q_{\rm BB} \in {\rm SO}^*(2n)/{\rm
  U} (n)$ \cite{so*} and $Q_{\rm FF} \in {\rm Sp}(2n) / {\rm U}(n)$,
which are symmetric spaces of type $D{\rm III}$ and $C{\rm I}$ --
hence the name $D{\rm III}|C{\rm I}$ for the present nonlinear
$\sigma$ model.  The same effective theory (restricted to the FF
sector due to the use of fermionic replicas) was obtained in
Ref.~\cite{mpafetal}, based on a quasiparticle Hamiltonian for a dirty
$d_{x^2-y^2}$ superconductor with orbital coupling to a magnetic
field.  This is no surprise, as that system belongs to symmetry class
$C$ and {\it the order parameter field $Q$ and its low-energy
  effective theory are determined solely by symmetry}.  (Incidentally,
the classification scheme of \cite{az_ns} assigns the quasiparticles
of the $d_{x^2-y^2}$ superconductor in zero field to class $C{\rm I}$.
According to \cite{mrz_rss}, the corresponding symmetric superspace is
$D|C$, also in agreement with the findings of Ref.~\cite{mpafetal}.)

In order to break parity and account for the Hall angle, one would
need to add to the Lagrangian a topological density proportional to
$\epsilon^{kl} {\rm STr} Q \partial_k Q \partial_l Q$, which is
closely related to Pruisken's $\theta$ term \cite{pruisken} well known
from the theory of the integer quantum Hall effect.  In two dimensions
this term integrates to a winding number and is nontrivial, since the
fundamental group of ${\rm U}(n)$ is $\Pi_1({\rm U}(n)) = {\bf Z}$ and
there exists the topological identity $\Pi_1({\rm U}(n) = \Pi_2({\rm
  Sp}(2n)/{\rm U}(n))$.  However, such a topological term does not
affect the results for the longitudinal spin and thermal
conductivities presented below and will therefore be omitted.

The maximum entropy derivation presented here does not supply
microscopic expressions for the couplings $\nu D_\parallel$ and $\nu
D_\perp$.  (We can express them in terms of the random matrix
parameters $w_{ij}$, but this is neither illuminating nor useful.)
These parameters can either be calculated from (quasi)classical
transport theory \cite{classic} or, better yet, taken from experiment.
In the latter case we extract the (bare) coupling constants from
experiments conducted at temperatures high enough so that the
transport is classical, and then {\it use the field theory
  (\ref{sigmod0},\ref{sigmod1}) to predict the quantum corrections
  that emerge at lower temperatures.}

\section{Weak localization in class $C$}

The field theory (\ref{sigmod0}) does not apply to charge transport, as
the condensate carries charge and quasiparticle charge is not a
constant of motion.  The energy, however, and for class $C$ also the
spin of a quasiparticle are conserved, which allows to probe for
quasiparticle transport and localization by measuring the {\it
  thermal} and {\it spin} transport.  To obtain the relevant transport
coefficients we start from the bilocal conductivity tensor
  $$
  \tau_{ll}({\bf x},{\bf x}';E) = \sum_{\alpha\alpha'} 
  v_{l\alpha}^{({\bf x})} G^{\rm R}_{\alpha\alpha'} ({\bf x},{\bf x'};E) 
  v_{l\alpha'}^{({\bf x'})} G^{\rm A}_{\alpha'\alpha}({\bf x'},{\bf x};E),
  $$
which describes the nonlocal linear response of the spin current to
a perturbation due to the Zeeman coupling with an applied field.  The
quantities $G^{\rm R}$ and $G^{\rm A}$ are the retarded and advanced
Gorkov Green functions and $v_{l\alpha} = \big( i (\sigma_3)_{\alpha
  \alpha} ({\buildrel \leftarrow \over {\partial_l}} - {\buildrel
  \rightarrow \over {\partial_l}}) - 2e A_l \big)/2m$ is the
$l$-component of the velocity operator.  We use the relation ${\cal H}
= - {\cal C} {\cal H}^{\rm T} {\cal C}^{-1}$ to express $G^{\rm A}$ at
energy $E$ by $G^{\rm R}$ at energy $-E$ \cite{bcsz}.  Disorder
averaging and the mapping on the nonlinear $\sigma$ model with $n = 2$
then turn the tensor $\tau_{ll}$ into a correlation function of the
conserved ${\rm OSp}(4|4)$ Noether current ${\cal J} = (Q \nabla
Q)^{{\rm B}11,{\rm B}22}$ of the field theory:
\begin{equation}
  \left\langle \tau_{ll}({\bf x},{\bf x}') \right\rangle = 
  (\pi\nu D_l)^2 \left\langle {\cal J}_l({\bf x}) 
    \bar{\cal J}_l({\bf x}') \right\rangle \;.
\label{correlator}
\end{equation}
The second superscript in the expression for the Noether current
refers to the pseudo charge, while the third distinguishes between the
two Green functions.  The symmetry breaking perturbation due to the
quasiparticle energy $E$ is incorporated into the formalism by setting
$\omega = {\rm diag} (E^+ , -E^-)$, where $E^\pm = E \pm i0$.

Next, let $\sigma_{ll} = (2\pi)^{-1} \int \left\langle \tau_{ll} ({\bf
    x} , {\bf x}^\prime ) \right \rangle d^3 x^\prime$ be the local
``spin'' conductivity for quasiparticles with fixed spin up or down.
To compute this quantity from the correlator (\ref{correlator}), we
adopt a rational parametrization for $Q$,
  $$
  Q = \pmatrix{1 &Z\cr \tilde Z &1\cr} \pmatrix{1 &0\cr 0 &-1\cr}
  \pmatrix{1 &Z\cr \tilde Z &1\cr}^{-1} \;.
  $$
Inserting this parametrization into the field theory action 
(\ref{sigmod0}), and doing the functional integral in Gaussian
approximation (tree level), we obtain $\sigma_{ll} = \sigma_{ll}^0$
with
  $$
  \sigma_{ll}^0 = \nu D_l \;,
  $$
which is the result expected from quasiclassical transport theory. 
The weak localization correction to $\langle \tau_{ll}({\bf x},{\bf
  x}^\prime) \rangle$ arises from one-loop graphs of the kind shown in
Fig.~1; see \cite{bcsz}.  The basic element of this graph is a
4-vertex representing the fourth-order term in the Taylor expansion of
$S$ with respect to $Z,\tilde Z$.  A double line oriented by an arrow
stands for the bare propagator $\langle Z \tilde Z \rangle_0$.  All
one-loop graphs are composed of three propagators and one
4-vertex.  Although these graphs appear as a calculational device for
organizing the field-theoretic perturbation expansion, they do have a
direct physical meaning, as follows.  Each of the two single lines in
Fig.~1 stands for a Feynman path contributing to the Gorkov Green
function $G^{\rm R} ({\bf x}, {\bf x}^\prime;\pm E)$.  Double lines
represent sums of impurity ladders with an arbitrary number of 
Andreev scattering events inserted.  It is seen from Fig.~1 that one
of the two Green function lines proceeds directly from the point ${\bf
  x}^\prime$ to the point ${\bf x}$, whereas the other one makes an
excursion in the form of a double loop.  The propagator associated
with the double loop is called the $D$-type cooperon \cite{az_ns}.
What is essential here is that the charge of the quasiparticle during
the second looping is exactly opposite to the charge during the first
looping.  This feature makes the $D$-type cooperon stable with respect
to disorder averaging irrespective of the orbital coupling to a
magnetic field, by canceling the Aharonov-Bohm phase $\oint {\bf A}
\cdot d{\bf l}$ accumulated in the loop.  Fig.~1 also indicates the
fact \cite{az_andreev} that the present variant of the weak
localization phenomenon already affects a single Green function and
thus the density of states.
\begin{figure} [h]
\leavevmode
\epsfxsize=6cm
\epsfysize=4cm
\centerline{\epsfbox{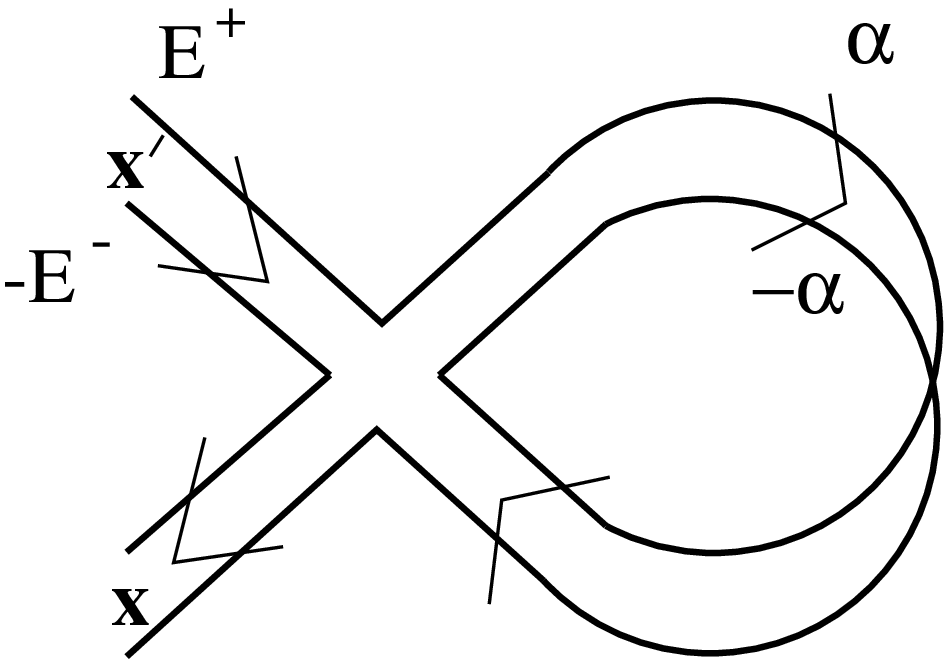}}
\caption{\label{FIG} One-loop diagram contributing to the correlator
$\langle \tau_{ll}({\bf x},{\bf x}^\prime) \rangle$.  The electric 
charge of the quasiparticle during the second looping $(-\alpha)$ is 
opposite to the charge during the first looping $(\alpha)$.}
\end{figure}

By evaluating the one-loop graphs in a similar manner as in
Ref.~\cite{bcsz}, we obtain
\begin{equation}
  \delta\sigma_{ll} = - {D_l \over \pi} {\rm Re} \int {d^3 k 
    \over (2\pi)^3} \left( D_\parallel^{\vphantom{2}} k_\parallel^2 
    + D_\perp^{\vphantom{2}} k_\perp^2 + 2iE \right)^{-1} \;.
\label{WL}
\end{equation}
The full spin conductivity is $\sigma_{ll} = \sigma_{ll}^0 + \delta
\sigma_{ll} + \ldots$.  Note that the correction is formally similar
\cite{lr} to that for class $A{\rm I}$, except that it explicitly
depends on energy.  In fact, it {\it disappears with increasing
  excitation energy}, or temperature, in agreement with the fact
\cite{bcsz} that moving up in energy causes crossover from class $C$
to class $A$, where weak localization is absent.  In dimension $d \le
2$ the integral over wave numbers is cut off in the infrared by the
inverse of the dephasing length $L_\varphi$ due to inelastic (or
quasielastic \cite{aag}) scattering, while for dimension $d \ge 2$ it
is UV-regularized by the inverse elastic mean free path.

Next recall that the quasiparticle spin is assumed to be conserved,
which allows to consider the sectors with spin up ($s = +1/2$) and
spin down ($s = -1/2$) separately.  Turning on the Zeeman coupling is
equivalent to shifting the excitation energy $E \to E - s \mu B$.  As
a result, the energy dependence of the weak localization correction
translates into a {\it field dependence}.  Note that this effect
differs from weak localization in disordered metals \cite{lr}, where
the orbital coupling to a magnetic field causes class $A{\rm I}$ to
cross over to class $A$.  In that case, the field scale is set by
$B_{\rm O} = (e D \tau_\varphi)^{-1}$ with $\tau_\varphi$ being the
dephasing time.  In the present case the relevant field scale is
$B_{\rm Z} = (\mu \tau_\varphi)^{-1}$.

To compute the thermal conductivity $\kappa$ at temperature $T$, we 
use the relation
  $$
  \kappa_{ll} = \sum_{s = \pm 1/2} \int_0^\infty \sigma_{ll}
  (E - s\mu B) \ {\partial f_T \over \partial T}(E) \ E \ dE \;,
  $$
where $f_T(E) = (1 + e^{E/T})^{-1}$ is the Fermi-Dirac distribution,
and our unit of temperature is such that $k_{\rm B} = 1$.  If the
energy dependence of $\sigma^0$ can be neglected in the range $0 < E
\lesssim T$, and if $T \lesssim {\rm Max}(\mu B , \Gamma_ \varphi)$,
{\it i.e.} $\delta\sigma$ is cut off by the Zeeman energy or the
dephasing rate $\Gamma_\varphi = \tau_\varphi^{-1}$, rather than by
the temperature, we may pull out $\sigma_{ll}(E)$ from under the
integral sign, thus obtaining an analog of the Wiedemann-Franz law:
\begin{equation}
  {\kappa_{ll} (B) \over  T} = {\pi^2\over 3} \sigma_{ll}(\mu B) \;.
  \label{WF}
\end{equation}
Here we have combined the spin up and spin down contributions, by
assuming the quasiclassical term $\sigma_{ll}^0$ to be unaffected by
the Zeeman splitting.

\section{Isolated vortices}

We now specialize to an extreme type-II $s$-wave superconductor in a
weak magnetic field (but well into the mixed state so that the field
is approximately homogeneous), where quasiparticles are bound to a
dilute array of vortex cores and the amplitude to hop from one vortex
to another is negligibly small.  In this case the problem reduces to a
set of decoupled one-dimensional theories, one for each vortex, and we
formally set $D_\perp = 0$.  The parameters of the one-dimensional
nonlinear $\sigma$ model were calculated by solving the Usadel
equation for a single vortex in \cite{bcsz}, where we found
$D_\parallel = C_2 v_{\rm F} \ell / 3 C_1$, and $\nu \int d^2 x_\perp
= 2\nu_{\rm N} \pi \xi^2 C_1$ if the integral extends over the area
occupied by one vortex.  The parameter $\xi$ is the dirty coherence
length, $\nu_{\rm N}$ is the density of states of the normal metal,
and $C_1 = 3.16$ and $C_2 = 1.20$ are numerical constants dependent on
the vortex profile.  Using the fact that the total number of vortices
equals the transverse area of the sample divided by half the square of
the magnetic length $l_B = \sqrt{2\pi / eB}$, we obtain
$\sigma_\parallel^0 = 4\pi C_2 \nu_{\rm N} (\xi / l_B)^2 v_{\rm F}
\ell / 3$ for the quasiclassical limit of the spin conductivity.  The
weak localization correction is given by
  $$
  \delta\sigma_\parallel = - {2 D_\parallel \over \pi l_B^2} 
  \ {\rm Re} \int {dk \over 2\pi} \left( D_\parallel k^2 + 
    \Gamma_\varphi + 2i(E - s\mu B) \right)^{-1} \;,
  $$
where inelastic events were incorporated by shifting the denominator
by $\Gamma_\varphi$.  This result applies when the dephasing length
$L_\varphi = \sqrt{D_\parallel / \Gamma_\varphi}$ is shorter than the
vortex length $L_\parallel$. In the opposite, mesoscopic regime
($L_\parallel \ll L_\varphi$) the weak localization effect was
worked out in \cite{bcsz}.

The low-temperature behavior of the thermal conductivity depends on
how $\Gamma_\varphi$ varies with $T$.  If we assume a power law
$\Gamma_\varphi \sim T^p$ with exponent $p < 1$ \cite{aag}, then
$\sigma_\parallel$ becomes constant in the energy range where
$df_T(E)/dT$ is essentially different from zero, and we get the
Wiedemann-Franz law (\ref{WF}) with
  $$
  \sigma_\parallel(\mu B) = \sigma_\parallel^0 - ( \pi l_{B}^2 )^{-1}
  {\rm Re} \sqrt{D_\parallel / ( \Gamma_\varphi + i\mu B)} \;.
  $$
In the high-field regime $\mu B \gtrsim \Gamma_\varphi$ the weak
localization correction to the thermal conductivity is cut off by the
Zeeman splitting, giving a characteristic dependence $\delta \kappa_
\parallel / T \sim - 1 / \sqrt{B}$.  On the other hand, if $p > 1$
then the relevant low-$T$ regime is $T \gg \Gamma_\varphi$, and the
weak localization effect is cut off by $T$ for low fields. In that
case, one finds $\kappa_\parallel / T = {\pi^2\over 3}\sigma^0 -
{3\over 4}\sqrt{\pi\over 2}(\sqrt{2}-1) \zeta(3/2) L_T / \pi l_B^2$,
{\it i.e.} the quantum correction is determined by the thermal length
$L_T = \sqrt{D_\parallel/T}$.

The above considerations apply to a vortex array in the dilute limit
near $H_{{\rm c}1}$.  As the field is increased, the quasiparticle
hopping rate between vortices in an $s$-wave superconductor grows
strongly.  When the field is tuned close to $H_{{\rm c}2}$, where the
system of vortex cores becomes dense, the diffusion constant $D_\perp$
gets large and the anisotropic field theory (\ref{sigmod0})
three-dimensional.  Since the quasiparticle states of the weakly
disordered three-dimensional system are extended, a delocalization
transition must take place with increasing field.  Note that this
transition is not in a new universality class, as the breaking of spin
rotation invariance by the Zeeman coupling reduces class $C$ to class
$A$ \cite{mpafetal}.  Nevertheless, the occurrence of such a
delocalization transition may be of experimental interest, for it can
be observed {\it by varying the magnetic field} (instead of the
disorder strength or the chemical potential).

\section{Weak Localization in the Cuprates}

We now adapt our results to the very interesting case of quasi
two-dimensional $d$-wave superconductors such as the cuprates.  As was
stated before, the low-energy quasiparticles of a dirty $d$-wave
superconductor in zero magnetic field belong to symmetry class $C{\rm
  I}$.  Weak localization effects in that class arise from two
distinct modes of quantum interference \cite{az_ns}: the cooperon of
type $A$, and the cooperon of type $D$.  The former is the natural
analog of the cooperon mode well known from the theory of disordered
metals \cite{lr}.  When time reversal symmetry is broken by a magnetic
field penetrating the superconductor, the $A$-type cooperon becomes
massive and disappears over a scale given by $B_{\rm O} = (e D \tau_
\varphi) ^{-1}$.  This crossover takes class $C{\rm I}$ into class
$C$, while leaving weak localization due to the $D$-type cooperon
intact.  As we have seen, the latter mode is cut off only by the
Zeeman energy, which becomes effective over the characteristic field
scale $B_{\rm Z} = (\mu \tau_\varphi)^{-1}$.  Using $\mu = 2\mu_{\rm
  B}= e / m$ and $D \sim k_{\bf F} \ell / m$ we see that the two
scales are separated by a large factor: $B_{\rm Z} / B_{\rm O} \sim
k_{\bf F} \ell$, {\it i.e.} the elimination of the $A$-type cooperon
by the orbital coupling to the magnetic field takes place at much
smaller fields than does the removal of the $D$-type cooperon by the
Zeeman energy.  This justifies our explicitly retaining the Zeeman
coupling, while burying the orbital coupling via the introduction of a
maximum entropy model.  In the following, we take the magnetic field
to be applied along the $c$-axis, and assume the system to be well
into the mixed state so that the field is approximately homogeneous.

The cuprates are highly anisotropic materials, consisting of weakly
coupled ${\rm Cu O}_2$ planes, for which $D_\parallel \equiv D_c \ll
D_{ab} \equiv D_\perp$.  At weak interlayer coupling, the continuum
approximation leading to (\ref{sigmod0}) is not justified in the
$c$-direction, and we need to restore the discrete layer structure.
This is done by making the replacement $D_\parallel ^{\vphantom{2}}
k_\parallel^2 \to 2 t_c ( 1 - \cos k_\parallel a )$, where $a$ is the
distance between layers and $t_c$ is the interlayer hopping rate.
Then, by performing the integral in (\ref{WL}) over the domain
$L_\varphi^{-1} < | k_\perp | < \ell^{-1}$ and $-\pi / a < k_\parallel
< \pi / a$, we obtain
\begin{eqnarray}
  \delta\sigma_{\perp} &=& - (2\pi^2 a)^{-1} {\rm Re} \ln
  \Big( F_s(\Gamma) / F_s(\Gamma_\varphi) \Big) \;,
  \label{WLcuprates}
  \\
  F_s(\varepsilon) &=& \sqrt{\varepsilon + 4t_c + 2i(E - s\mu B)}
  + \sqrt{\varepsilon + 2i(E - s\mu B)} \nonumber \;,
\end{eqnarray}
where $\Gamma = D_\perp / \ell^2$ and $\Gamma_\varphi = D_\perp / 
L_\varphi^2$ are the elastic and inelastic scattering rates.

To evaluate the consequences of this general formula, one needs to
distinguish cases.  For brevity, we concentrate on the limit defined
by the condition that elastic scattering sets the largest energy
scale: $\Gamma \gg {\rm Max} (4t_c , 2E, \mu B)$.  Consider first the
case $4t_c \lesssim {\rm Max}(2E,\mu B)$, which physically means that
the coherence of the quantum interference modes is destroyed before
quasiparticles have a chance to hop between layers.  The layers then
effectively decouple, yielding a two-dimensional system, and the
formula for $\delta \sigma_\perp$ becomes
  $$
  \delta\sigma_\perp = - (4\pi^2 a)^{-1} {\rm Re} \ln \Big( 
    \Gamma / (\Gamma_\varphi + 2iE - 2is\mu B) \Big) \;.
  $$
The appearance of a logarithm is characteristic of weak localization
in two dimensions.  For the in-plane thermal conductivity we get
  $$
  {\delta\kappa_\perp(B) / T} = - (12 a)^{-1} \ 
  {\rm Re} \ln \Big( \Gamma / (\Gamma_\varphi + i\mu B) \Big) \;,
  $$
provided that the conditions of validity of the Wiede\-mann-Franz law 
(\ref{WF}) are satisfied.  Note that in contrast with
three-dimensional metals, where weak localization is a rather minute
effect, the correction here can easily exceed $10\%$ under
experimental conditions.  This is because the relative size
$\delta\kappa / \kappa$ is roughly given by the inverse of the
dimensionless intralayer coupling constant $2\pi \nu_{2d} D_\perp$,
whose value in zero field has been estimated \cite{palee,mpafetal} to
be not much in excess of unity.

In the opposite limit, where $4 t_c$ is much larger than $\mu B$ and
$\Gamma_\varphi$, but still smaller than $\Gamma$, the field
dependence of the weak localization correction to the thermal
conductivity becomes three-dimensional:
  $$
  {\delta\kappa_\perp(B) \over T} = - {1 \over 6a} \left( \ln 
    \sqrt{\Gamma / t_c} - {\rm Re} \sqrt{( \Gamma_\varphi 
      + i\mu B) / 4 t_c } \right) \;,
  $$
where again the law (\ref{WF}) was assumed.  Note that the above 
expressions for $\delta \kappa_\perp (B)$ {\it increase} with $B$.

To summarize, weak localization in class $C$, cut off by the Zeeman
splitting, causes the thermal conductivity to increase with the
magnetic field at sufficiently low temperatures.  To make this more
quantitative, we need to specify the field/temperature range where the
effect becomes observable.  The answer is provided by the value of the
spin magnetic moment of the electron (with a $g$-factor of 2), which
is 1.35 K/T in suitable units.  As a result, if the field strength is
of the order of 1 Tesla, the weak localization induced field
dependence sets in at temperatures below one 1 Kelvin (unless for some
unexpected reason the dephasing rate $\Gamma_\varphi$ is anomalously
large).

We now wish to elucidate whether such an effect might already be
visible in recent experiments.  The discussion is somewhat complicated
by an ongoing debate concerning the leading, quasiclassical term
$\kappa^0$.  Let us summarize the current situation as we see it.

Krishana {\it et al.} \cite{krishana} measured the magnetic field
dependence of the thermal conductivity in a BSCCO system for
temperatures $T \geq 6 \ {\rm K}$.  After an initial decrease at weak
fields, they observed a sharp kink at $B^* \sim \sqrt{T}$, followed by
a wide plateau for $B > B^*$.  The nonanalyticity at $B^*$ has been
interpreted \cite{laughlin} as a phase transition to a new ground
state with a secondary $id_{xy}$ order parameter.  We will not be
particularly concerned with that issue here.  (The addition of an
$id_{xy}$ component to the order parameter is fully compatible with
the symmetries of class $C$ and, if disorder is present, the field
theory (\ref{sigmod0}) for the quasiparticle excitations remains
qualitatively unchanged.)  From the observation of field independence
over a sizable range of temperatures, one deduces \cite{krishana} that
both the electronic and the phonon contribution to the thermal
conductivity must be individually constant.  The constancy of the
electronic part was initially attributed to the $d_{x^2-y^2} +
id_{xy}$ state being fully gapped, {\it i.e.} to the complete absence
of low-energy quasiparticle excitations.  This explanation has been
challenged by experimental data of Aubin {\it et al.} \cite{aubin}.
While confirming the results of Ref.~\cite{krishana} for $T > 5 \ {\rm
  K}$, these data reveal the emergence of a positive thermal
magnetoconductance at lower temperatures $T \lesssim 1 \ {\rm K}$.
(The data also show pronounced hysteresis effects whose interpretation
remains controversial.)  Taking the constancy of the phonon
contribution for granted, the observation of such dependence strongly
indicates a residual density of quasiparticle states at zero energy.
The existence of such states is no surprise.  Indeed, in the mixed
state of a superconductor with $d_{x^2-y^2}$ wave symmetry a residual
density of states is expected even in the absence of disorder, because
some fraction of the low-energy quasiparticles (close in momentum to
the $d$-wave nodes) are Doppler shifted to zero energy by the
supercurrent circulating around the vortices (the Volovik effect
\cite{volovik}), which leads to $\nu(E=0,B) = \nu_{\rm N} \sqrt{B /
  B_{{\rm c}2}}$.  (For a recent discussion of the same effect for a
ground state with $d_{x^2-y^2} + id_{xy}$ symmetry, see \cite{mao}.)
The residual density of states created by this mechanism is
approximately {\it constant in energy} below the average Doppler shift
scale $E_B$, roughly estimated by $E_B / T_{\rm c} \simeq \sqrt{B /
  B_{{\rm c} 2} }$.  Disorder can only broaden the range of energy
independence of $\nu$.  Hence, assuming $B_{\rm c2} \sim 100$ T and a
superconducting transition temperature $T_{\rm c} \sim 100$ K, the
energy scale $E_B$ is of order 10 K for fields of a magnitude of about
1 T.

Now recall the experimental observations reported in \cite{aubin}: an
electronic thermal conductivity which is independent of the magnetic
field for $T \gtrsim 5$ K (and $H > H^*$), and begins to increase with
$B$ below $T \simeq 1$ K.  (According to a footnote in \cite{aubin},
the same effect has been seen in YBCO.)  The first point to address is
the field independence at the higher temperatures.  Franz \cite{franz}
has recently proposed a model for the quasiclassical thermal
conductivity $\kappa^0$, in which the increase of $\nu$ with the field
is exactly canceled by a concomitant decrease of the quasiparticle
mean free path $\ell$.  The model assumes scattering from the
superflow due to randomly positioned vortices.  In contrast, another
recent theory \cite{kuebert} argues in favor of the dominant
scattering mechanism being impurities close to the unitarity limit.
We will not pursue here the discussion as to which is the correct
model to use.  With the microscopic theory of the plateau effect being
a subject of debate, our philosophy is to accept it as an {\it
  experimental fact} that the field variation of $\nu$ and $\ell$ is
such as to cancel in $\kappa^0 \sim \nu(B) \ell(B)$.  The question to
address, then, is {\it why a field dependence sets in when the
  temperature is lowered}.  We argue that this is at least in part due
to weak localization.  As we have seen, weak localization in the mixed
state of dirty $d$-wave superconductors is a phenomenon on safe
theoretical ground, is sizable in magnitude, and is expected to occur
at the right temperature and field scales to match the experiment
\cite{aubin}.  To preclude any confusion, we stress that the effect
under consideration is distinct from weak localization in combination
with Aslamasov-Larkin fluctuations, which have been invoked in
\cite{ong} to explain the negative thermal magnetoconductance observed
in a dirty LSCO system at much higher temperatures.

Theories proposed by previous authors attribute the temperature
variation of $d\kappa(B,T) / dB$ to the leading (quasiclassical) term,
$\kappa^0 \sim \nu\ell$.  Given the low-energy constancy of the
density of states, such a variation would have to arise from an energy
(or temperature) dependence of the elastic mean free path.  Possible
explanations are: i) low-energy transparency of $d$-wave vortices to
quasiparticles \cite{franz}, and ii) energy-dependence of the elastic
scattering rate due to impurities near the unitarity limit
\cite{kuebert}.  The challenge to these scenarios is to explain why
for fields $B \sim 1$ T the effect sets in at temperatures around 1 K.
In the weak localization scenario we have described, this comes about
very naturally if $\Gamma_\varphi$ is determined by thermal
broadening, since $\mu = 1.35 \ {\rm K/T}$.

A clear difference is that weak localization effects {\it continue to
  be enhanced} with decreasing temperature -- they ultimately drive
the system to an insulator by localization of all quasiparticle states
-- whereas the energy dependence of the elastic mean free path
saturates.  To discriminate, it is therefore desirable to push the
experimental measurements to the lowest temperatures possible.  In
order to achieve a quantitative description based on formula
(\ref{WLcuprates}), it will be necessary to take the field dependence
of $\ell$ into account.  Our suggestion is to extract the density of
states $\nu(B)$ from measurements of the specific heat, and then
deduce $\ell(B)$ from the quasiclassical formula $\kappa(B) \sim
\nu(B) \ell(B)$, valid at high temperatures ($T \gtrsim 5$ K).  As far
as the temperature dependence of the dephasing rate $\Gamma_\varphi$
is concerned, a phenomenological model needs to be used.  To our
knowledge, a theory for this quantity in the mixed state of dirty
$d$-wave superconductors does not exist.  In the long run, weak
localization may turn out to be the appropriate tool to {\it measure}
$\Gamma_{\varphi}$, as is established practice in disordered metals
\cite{lr,aag}.

\section{Conclusion}

Noninteracting electrons subject to disorder and a magnetic field are
well known to belong to the standard universality class $A$ (unitary
symmetry, $\beta = 2$).  When spin-singlet pairing correlations are
added, the universality class of the low-energy quasiparticles changes
to type $C$.  It has been shown that the transport properties of these
quasiparticles are unconventional.  In particular, there exist modes
of destructive quantum interference which survive the orbital coupling
to a magnetic field.  They are cut off at higher fields by the Zeeman
coupling, thereby giving rise to a field dependent quantum (or weak
localization) correction to the low temperature thermal conductivity,
with the characteristic scale given by $\mu = 1.35$ K/T.  A good place
to look for such corrections experimentally are disordered
low-dimensional superconductors, such as the cuprates, in the mixed
state.

On general symmetry grounds, the low-energy effective field theory for
quasiparticles in class $C$ is predicted to be a nonlinear $\sigma$
model of type $D{\rm III}|C{\rm I}$.  The Lagrangian of this field
theory has a universal form, independent of the symmetry of the order
parameter ($s$, $d$, {\it etc.}), as long as the superconductor
conserves the quasiparticle spin and is penetrated by magnetic flux.
The role of the superconducting ground state is merely to determine
the values of the field theory coupling constants, their anisotropy,
and their dependence on energy and magnetic field.  Quantitative
predictions for the weak localization corrections to transport can be
made once the values of the couplings and their dependences have been
obtained, either from quasiclassical transport theory or from
experiment.  We advocate the use of such predictions in understanding
the low-temperature experiments of Aubin {\it et al.} \smallskip

{\bf Acknowledgment.} This research was supported in part by the
Deutsche Forschungsgemeinschaft, SFB 341 (K\"oln-Aachen-J\"ulich).
One of the authors (M.R.Z.) thanks A. Freimuth for a discussion.


\begin{references}
\bibitem{bahcall} This symmetry, which is vital for the weak
  localization correction of class $C$, is missing from the random
  matrix model proposed by S.R. Bahcall, Phys. Rev.  Lett. {\bf 77},
  5276 (1997).
\bibitem{so*} The group ${\rm SO}^*(2n)$ is a noncompact version of
  the orthogonal group ${\rm SO}(2n)$.
\end{references}
\end{document}